\documentclass{emulateapj}
\usepackage{apjfonts}
\usepackage{graphicx}
\usepackage{natbib}
\usepackage{amsmath}
\usepackage{amssymb}

\slugcomment{Submitted to Astrophysical Journal}

\begin{document}



\title{A New Probe of the Distribution of Dark Matter in Galaxies}

\author{Sukanya Chakrabarti \altaffilmark{1}}
\altaffiltext{1}
{777 Glades Road, Physics Dept., Florida Atlantic University, Boca Raton, FL 33431; schakra1@fau.edu}

\begin{abstract}
The scale radius of dark matter halos is a critical parameter for specifying the density distribution of dark matter, and is therefore a fundamental parameter for modeling galaxies.  We develop here a novel, observationally motivated probe to quantitatively infer its value.  We demonstrate that disturbances in the extended atomic hydrogen gas disks of galaxies can be used to infer the scale radius of dark matter halos.  Our primary metric is the phase of the $m=1$ mode of the disturbance in the outskirts of the gas disk, which we take to be produced by a tidal interaction.  We apply the method to the Whirlpool Galaxy, which has an optically visible satellite.  We explore potential degeneracies due to orbital inclination and initial conditions and find our results to be relatively insensitive to these considerations.   Our method of tracing the dark potential well through observed disturbances in outer gas disks is complementary to gravitational lensing.
\end{abstract}


\section{Introduction}

The cold dark matter (CDM) paradigm of structure formation is successful at recovering the basic skeletal structure of the universe -- the large-scale distribution of galaxies (Springel et al. 2006).  However, the agreement between theory and observation is less secure when this model is applied to galactic (and sub-galactic) scales.   While the existence of dark matter halos in galaxies was observationally inferred in the 1970's (Rubin et al. 1977; Bosma 1978), there are few observational probes that can be used to infer the details of the distribution of dark matter.  Simulations have not yet been able to resolve the mitigating effects of gas cooling (Gnedin et al. 2004) versus outflows (Governato et al. 2010), which can act to increase (decrease) the distribution of dark matter in galaxies.  In the coming decades, the CDM paradigm will be increasingly vetted against detailed observations of the innards of galaxies that will delve even more deeply into galaxy evolution.  

Navarro, Frenk \& White (1996; 1997; NFW) found that dark matter halos formed through dissipationless hierarchical clustering have an universal density profile.  Due to the absence of scale in this heirarchical structure formation scenario, dark matter haloes are scaled versions of each other, with the total mass being the scaling parameter.  The NFW density profile is given by:
\begin{equation}
\rho(r) = \frac{\delta_{c}\rho_{c}}{(r/R_{s})\left(1+r/R_{s}\right)^{2}}
\end{equation}
where $\rho_{c}$ is the critical density for closure of the universe, and $\delta_{c}$ is a characteristic overdensity for the halo.  The scale radius, $R_{s}=R_{200}/c$ is a characteristic radius, and $c$ is the concentration parameter that relates this characteristic radius and $R_{200}$, which is the radius at which the mean enclosed dark matter density is 200 times the critical density.   The characteristic overdensity of the halo, $\delta_{c}$, is simply a function of the concentration parameter $c$ (NFW).  The NFW profile differs from the Hernquist (1990) profile in its asympotic behavior, scaling as $r^{-3}$ rather than $r^{-4}$ for $r/r_{s} \gg 1$.  The scale radius of the dark matter halo along with the concentration parameter specify the density distribution for a given halo mass in the NFW formalism.  Currently, these parameters are prescribed from large cosmological N-body simulations (Bullock et al. 2001; Maccio et al. 2008).  It is not presently clear how we can determine these values for individual galaxies.  What we seek to do here is to develop an independent, observationally motivated probe of the scale radius of dark matter halos, and apply it to the well-known Whirlpool Galaxy (M51) to infer its scale radius.  The  density switches its behavior at the scale radius, from $r^{-1}$ for $r \ll R_{s}$ to $r^{-3}$ for $r \gg R_{s}$, and passes through the transitional $r^{-2}$ region for $r \sim R_{s}$.  Thus, the scale radius of dark matter halos is a critical parameter for specifying the density distribution of dark matter in galaxies, and therefore a fundamental parameter for modeling galaxies.  

The extended atomic hydrogen (HI) disks of galaxies provide an unique probe of galaxy evolution.  They are ideal tracers of tidal interactions with satellites and the galactic gravitational potential well.  We recently developed a novel method whereby one can infer the mass, and relative position (in radius and azimuth) of satellites from analysis of observed disturbances in outer gas disks, without requiring knowledge of their optical light (Chakrabarti \& Blitz 2009, henceforth CB09; Chakrabarti \& Blitz 2011, henceforth CB11; Chakrabarti, Bigiel, Chang \& Blitz 2011, henceforth CBCB; Chang \& Chakrabarti 2011, henceforth CC11).   We applied this method to M51 and inferred that its satellite has a mass  one-third that of the primary galaxy, with a pericentric approach distance of 15$~\rm kpc$.  We found these estimates to be corroborated by observations (Smith et al. 1990) and recent simulation studies (Salo \& Laurikainen 2000; Dobbs et al. 2010).  Moreover, at the time when our simulations achieve the best-fit to the HI data, the azimuth of the satellite in the simulations agrees closely with its observed location.  CBCB note that the derivation of these numbers is uncertain at the factor of two level due to variations in the initial conditions of the simulated M51 galaxy, orbital inclination and orbital velocity of the satellite.  We call this method "Tidal Analysis".

We frame the problem here in the following way -- for galaxies like M51, for which we have demonstrated that our Tidal Analysis method works, can we determine the scale radius of the dark matter halo of M51? We take the satellite mass and pericentric distance from our earlier studies as an input.  It is worth noting that such a procedure has viable and immediate applicability.  The THINGS survey produced HI maps of local spiral galaxies (Walter et al. 2008), many of which display disturbances in the outskirts (Bigiel et al. 2010).   Attempting to constrain the dark matter halo in this way is similar in spirit to earlier work on analysis of stellar tidal tails by Mihos, Dubinski \& Hernquist (1998).  If the satellite mass and pericentric distance of the satellites that produce these disturbances can be characterized, as for M51 (CBCB), then as we show below, the phase of the $m=1$ mode directly yields the scale radius of the dark matter halo.  This method of probing the dark matter mass distribution is independent of the stellar light and is complementary to strong gravitational lensing, which primarily probes regions interior to the Einstein radius, $\sim 10~\rm kpc$ in spirals (Wright \& Brainerd 2000; Treu \& Koopmans 2002).  The weak lensing signal has also been exploited by stacking the surface density contrast to produce mean density profiles for clusters (Sheldon et al. 2008) and total galaxy masses (Mandelbaum et al. 2006).  This paper is organized as follows:  in \S 2, we demonstrate that the phase of the $m=1$ mode can be used to quantitatively infer the scale radius of dark matter halos and apply the method to M51.  We discuss caveats and future work in \S 3, and conclude in \S 4.

\section{Results}

The simulations we discuss here have the same setup as in CBCB.  Since that paper discusses the simulation setup in detail, we only briefly discuss it here.  We carry out SPH simulations using the GADGET code (Springel 2005), of M51 interacting with its companion.  CBCB carried out a simulation parameter survey and compared the resultant Fourier amplitudes of the low order modes of the gas surface density with the observed HI data of M51.  They found that placing simulations on a variance vs variance plot (where the variance is  with respect to the low order modes of the simulations and the data) made the best-fit simulations visually apparent.  Specifically, CBCB found that the best-fit to the HI data occurred for a 1:3 mass ratio satellite with a pericentric distance of $15~\rm kpc$, parameters that are corroborated observationally and are in agreement with other simulation studies (Smith et al. 1990; Dobbs et al. 2010; Salo \& Laurikainen 2000).  CBCB also found that the azimuthal location of the companion of the best-fit simulation agrees very closely with the observed location of M51's companion, at the time when the Fourier amplitudes most closely match the data.  Thus, CBCB concluded that analysis of observed disturbances in the extended HI disks of galaxies can be used to infer the mass and current distance (in radius and azimuth) of galactic satellites.  Earlier work in this series of papers presented the basic reasoning as to why the mass-pericentric approach degeneracy in the tidal force can be broken when the time integrated response of the primary galaxy is considered (CB09), and described the method to find the azimuth of galactic satellites from the phase of the modes (CB11).

Our main goal in this paper is to determine whether the scale radius of the dark matter halo in the primary galaxy can be inferred from analysis of observed disturbances in the extended HI disk of M51.  Since we have earlier characterized the mass and pericentric approach distance of M51's companion, we take these quantities as inputs in our study here.  Here, we primarily vary the density profile (and hence the potential depth) of the dark matter halo of M51, which we take to follow an NFW profile.  We investigate whether varying the scale radius of the dark matter halo of M51 will be reflected in the disturbances of the extended HI disk cleanly enough to allow us to infer its value.

 We begin by studying the gas density response as M51 interacts with a 1:3 mass ratio satellite with a pericentric distance of $15~\rm kpc$ (parameters we derived in CBCB), while we vary the scale radius of the dark matter halo.  We vary the scale radius from low to high values ($R_{s}=11-32~\rm kpc$) to investigate its effect on the resultant disturbances in the HI disk.  The scale radius is related to the concentration parameter and the outer radius of the dark matter halo, i.e., $R_{s}=R_{200}/c$.  Therefore, we can either hold $R_{200}$ constant and vary the concentration parameter, or hold the concentration parameter constant and vary $R_{200}$.  The mass of the dark matter halo scales as $R_{200}^{3}$ ($M_{200}=200\rho_{c}\frac{4}{3}\pi R_{200}^{3}$), while the concentration parameter is related to the angular momentum of the halo as motivated by the Mo, Mao \& White (1998) formalism (Springel et al. 2005), and affects the size of the baryonic disk.  Our earlier models (CBCB) were based on the $R_{200}=160~\rm h^{-1}~kpc$ case, which gives a galaxy mass consistent with observational estimates (Leroy et al. 2008).  We first  set $R_{200}=160~\rm h^{-1}~kpc$ and vary the concentration parameter, which is equivalent to varying the scale radius of the dark matter halo.  Below we show that varying the concentration parameter and $R_{200}$ such that the scale radius is constant gives nearly identical results for the phase of the $m=1$ mode in the outskirts, which demonstrates that the critical parameter that governs the formation of these disturbances (once we have an observational handle on the mass of the galaxy from the rotation curve) is the scale radius.  Thus, the two cases we focus on here are:  1) holding $R_{200}$ constant and varying $c$ which corresponds to varying $R_{s}$; this means we hold the mass of the dark halo constant while we vary the scale radius, where the density switches from $r^{-1}$ for $r \ll R_{s}$ to $r^{-3}$ for $r \gg R_{s}$, and 2) holding $R_{s}$ constant and varying $R_{200}$ (which will then also vary $c$) and therefore the mass of the dark halo.
 
\begin{figure}[hb]
\begin{center}
\includegraphics[scale=0.7]{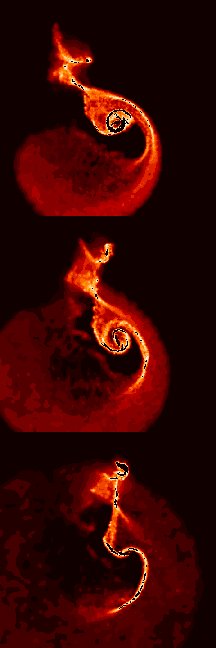}
\caption{A series of controlled experiments showing how M51's gas density response varies as the scale radius of the dark matter halo is varied.  We hold $R_{200}$ constant here as we vary the concentration parameter (scale radius) from large (low) to high (small) values.  The mass of the satellite and pericentric approach are taken from CBCB.  From top to bottom the cases are: (a) $R_{s}=11$, c= 14, (b) $R_{s}=17$, c = 9.4, the standard case, and (c) $R_{s}=32$, c = 5.  \label{f:m51CVar}}
\end{center}
\end{figure}

Figure \ref{f:m51CVar} shows the resultant gas density response of M51 when we vary the scale radius (concentration) of the dark matter halo of M51 from a small (large) value ($R_{s}=11,c=14$), to the fiducial value used earlier by CBCB ($R_{s}=17,c=9.4$), to a low (high) value ($R_{s}=32,c=5$).  Varying the scale radius varies the potential depth of M51 and will therefore dramatically impact the formation of tidal features, as we see clearly from Figure \ref{f:m51CVar}.   Steeper potential wells, which are produced by steeper density profiles, are more effective at holding on to the gas (Mihos, Dubinski \& Hernquist 1998).  Therefore, the $R_{s}=11,c=14$ case yields more tightly wound structures relative to the fiducial value of the scale radius ($R_{s}=17,c=9.4$), as well as of course the largest scale radius we consider here ($R_{s}=32,c=5$), where the density profile follows the shallow slope of $r^{-1}$ nearly all the way out to where we can probe the extended HI disk of M51.   

\begin{figure}[ht]
\begin{center}
\includegraphics[scale=0.4]{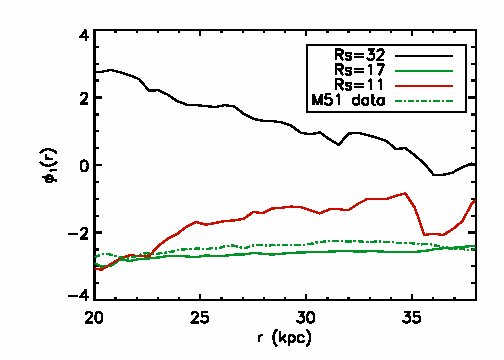}
\includegraphics[scale=0.4]{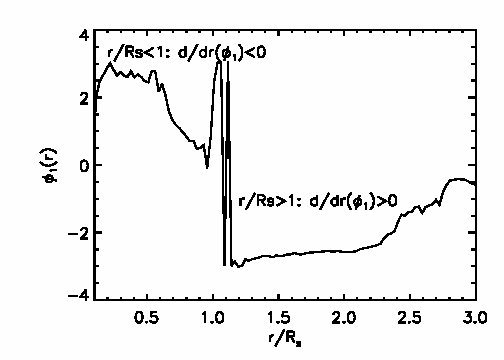}
\caption{(a) Phase of the $m=1$ mode as the scale radius of the dark matter halo is varied, shown at the same times as the gas density images.  The three cases correspond to probing regions inside the scale radius (the low concentration case), close to the scale radius (the standard C=9.4 case), and outside the scale radius (a high concentration case).  (b) Phase of the $m=1$ mode as a function of $r/R_{s}$ for the $R_{s}=17$ case.  Marked are the three regimes of interest.  $d/dr(\phi_{1})<0$ for $r/R_{s} < 1$, transitions at $r /R_{s}\sim 1$, and $d/dr(\phi_{1})>0$ for $r/R_{s} >1$ until the edge of the gas disk.  $R_{200}$ is held fixed here as the concentration is varied.  \label{f:m51m1phaseCvar}}
\end{center}
\end{figure}

Figure \ref{f:m51m1phaseCvar}(a) shows how the phase of the $m=1$ mode varies as we vary the scale radius of the dark matter halo in the three cases shown in Figure \ref{f:m51CVar}.  For projected gas surface density denoted $\Sigma(r,\phi)$, we calculate the phase of individual modes ``m'' by taking the FFT, where below we have set $m=1$:
\begin{equation}
\phi(r,m) = \arctan\frac{\left[-Imag~FFT~\Sigma(r,\phi) \right]}{\left[-Re~FFT~\Sigma(r,\phi)\right]} \; .
\end{equation}
The phase of the modes contains information on the shape of the spiral planform, i.e., tightly wrapped spirals will have a sharp gradient in the phase, while open spirals will have a flatter profile (Shu 1983; Chakrabarti \& Blitz 2011).  For the fiducial value of $R_{s}=17 (c=9.4)$ used by CBCB (shown in the solid green line), the phase of the $m=1$ mode in the simulation is fairly flat, as is that of the HI data of M51, shown in the dash-dotted green line.  However, for the other cases, i.e., $R_{s}=11$ and $R_{s}=32$,  the phase is either steeply falling or rising.  For the $R_{s}=32~\rm kpc$ case, the density profile follows a shallow $r^{-1}$ slope nearly all the way to the edge of the HI disk of M51, and as such this is the loosest spiral (nearly a straight line) produced.  The other extreme is the $R_{s}=11~\rm kpc$ case where the density follows a $r^{-3}$ profile in the radial range shown in Figure \ref{f:m51m1phaseCvar}(a).  Since the density profile is much steeper for the latter case, it is harder to "unwrap" the spiral, leading to a tighter spiral planform and positive gradient in the phase of the $m=1$ mode.  The radial variation of the phase, therefore, gives us a handle on the scale radius.  

There are three radial regions of interest in the behavior of the phase, and we mark them explicitly in Figure \ref{f:m51m1phaseCvar}(b) for the $R_{s}=17~\rm kpc$ case.  Here, we plot the phase as a function of the dimensionless variable $r/R_{s}$.  We see that $d/dr(\phi_{1}) < 0$ when $r/R_{s} < 1$, there is a transition region for $r/R_{s} \sim 1$, and $d/dr(\phi_{1}) > 0$ when $r/R_{s} > 1$, until the edge of the gas disk is reached.  In \S 2.1, we show that the (slight) uncertainty in the phase that arises due to variations in initial conditions and orbits does not preclude us from differentiating between the cases shown in Figure \ref{f:m51m1phaseCvar}.  If one can observe the transitional region, where the gradient in the phase switches from declining to increasing, then one can infer the value of the scale radius.  The shape of the phase of the $R_{s}=17~\rm kpc$ case is quite flat, as is that of the HI data for M51.  This comparison therefore indicates that the scale radius of the dark matter halo in the Whirlpool Galaxy is $\sim 17~\rm kpc$.

Our inference of a scale radius of $17~\rm kpc$ for M51 is comparable to values expected from dissipationless cosmological simulations (Bullock et al. 2001).  These simulations predict a 68 \% spread in concentration values from 9.2-21.3 for virial mass halos of $\sim 10^{12} M_{\odot}$ having virial radii $R_{\rm vir} \sim 200~\rm h^{-1}~\rm kpc$, which would yield $R_{s} \sim 13-31~\rm kpc$.  More recent work (Maccio et al. 2008) finds similar concentration-mass relations.  The NFW profile emerges in simulations that do not include baryonic physics.  One may well expect that baryonic processes, such as gas cooling or supernova feedback (Gnedin et al. 2004; Governato et al. 2010) would alter the density profile of the dark matter halo.  It is reassuring therefore that derivation of the scale radius for a specific local spiral with this new method (that relies on an analysis of disturbances in the extended gas disk) yields a value that is consistent with the expected range from dissipationless, cosmological simulations.

\begin{figure}[hb]
\begin{center}
\includegraphics[scale=0.5]{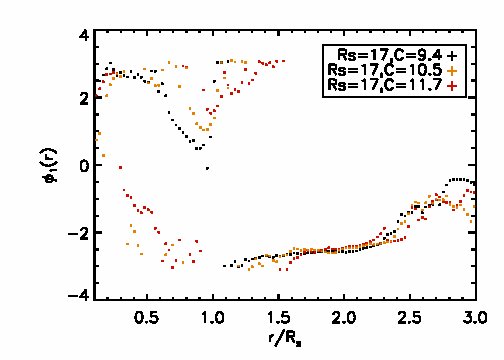}
\caption{The variation of the phase as the concentration is varied (by varying $R_{200}$) while holding the scale radius of the dark matter halo constant.   The black symbols mark the $C=9.4,R_{s}=17$ case, the orange marks the $C=10.5,R_{s}=17$, and the red the $C=11.7,R_{s}=17$.  As is clear, the behavior of all three is nearly the same for $r/R_{s} > 1$, and all three have similar slopes in the three regimes of interest noted in Figure \ref{f:m51m1phaseCvar}.   For $r/R_{s} < 1$, the different concentration parameters yield different disk sizes, which is why the amplitude of the phase is different in the inner regions.  \label{f:m51CVarRs17}}
\end{center}
\end{figure}

It is important to emphasize that for $r/R_{s} > 1$, our results depend essentially on the scale radius of the dark matter halo, and not on the concentration parameter.  We demonstrate this in Figure \ref{f:m51CVarRs17}, where we plot the phase of the $m=1$ mode of three cases where we vary the concentration parameter and $R_{200}$, but hold the scale radius fixed.  As is clear, for $r/R_{s} > 1$, the phase depends primarily on the scale radius (until we reach the edge of the gas disk; the size of the gas disk does depend on the concentration parameter, which is why there are small differences at the very edge).  All three cases, for fixed $R_{s}$, have the same type of slope in the three regimes of interest, i.e., for $r/R_{s} < 1$, $d/dr(\phi_{1}) < 0$, there is a transition region for $r/R_{s} \sim 1$, and $d/dr(\phi_{1}) > 0$ when $r/R_{s} > 1$.  For $r/R_{s} < 1$, the concentration parameter affects the size of the disk, and hence dramatically affects the amplitude of the phase in the inner regions.

Ideally, one would observe the transitional region where the phase of the $m=1$ mode switches from $d/dr(\phi_{1}) < 0$ for $r/R_{s} < 1$ to $d/dr(\phi_{1}) > 0$ for $r/R_{s} > 1$ to infer the value of the scale radius for a specific tidally interacting galaxy.  If $R_{s}$ is so large that it is comparable to the extent of the HI disk ($R_{HI}$), then one can at best put a lower limit on the scale radius, i.e., if we observe that the phase of the $m=1$ mode continues to have a negative gradient out to $R_{HI}$, then $R_{s}$ is greater than $R_{HI}$.  HI disks are generally more extended than the optical radius (Wong \& Blitz 2002), so the simulation modeling of the baryonic regions can in general be informed by optical or near-infrared photometry to yield stellar masses and disk sizes.  

\subsection{Dependence on Initial Conditions}

\begin{figure}[hb]
\begin{center}
\includegraphics[scale=0.5]{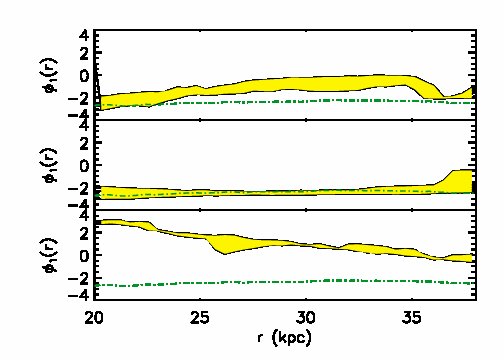}
\caption{Phase of the $m=1$ mode as the scale radius of the dark matter halo is varied from (a) $R_{s}=11$, c=14, (b) $R_{s}=17$, c = 9.4, the standard case, and (c) $R_{s}=32$, c= 5, shown at the same times as the gas density images.   Each figure displays the range in the phase as initial conditions (specifically, the gas and bulge fraction of the simulated M51 galaxy and orbital inclination of the satellite) are varied for a given concentration-scale radius value.  The phase of the data of M51 is overplotted in each figure as a dot-dashed green line.  \label{f:m51m1phaseICsvar}}
\end{center}
\end{figure}

We have investigated the dependence of the phase of the $m=1$ mode at the best-fit time on parameters aside from the scale radius.  Figure \ref{f:m51m1phaseICsvar} (a-c) depicts the variation (the spread is shaded in yellow) of the phase of the $m=1$ mode for a given concentration-scale radius value as the bulge and gas  fraction of the simulated M51 galaxy are varied, along with the orbital inclination of the satellite, for the three scale radius-concentration cases we have considered, namely (a) $R_{s}=11,C=14$, to (b) $R_{s}=17, C=9.4$, to (c) $R_{s}=32, C=5$.  Specifically, we consider here the fiducial model from CBCB that does not include a bulge, as well as a bulge mass fraction $m_{b}=M_{\rm bulge}/M_{\rm total}=0.012$.  We consider gas fractions $f_{g}=m_{\rm gas}/m_{\rm disk}$ (where $m_{\rm disk}$ is the disk mass) of $f_{g}=0.142$ (the fiducial case) and $f_{g}=0.25$.  We consider the fiducial case from CBCB of a co-planar orbit for M51's satellite, as well as an inclined orbit (the "e" orbit as denoted in CBCB).  

As is clear, while varying initial conditions has some effect on the phase, it does not change the essential character of the phase to the point that it would preclude us from discriminating from the different scale radius cases.  In each panel, the phase of the $m=1$ mode of the HI data of M51 is overplotted as a dot-dashed green line for comparison.  We found earlier in CBCB that varying the equation of state of the gas in the primary galaxy, and modestly varying the orbital velocity of the satellite,  does not significantly affect our results, and we do not vary those parameters here.  In the future, we will carry out a larger parameter survey and study the detailed effects of these and other parameters on this metric.  Our initial study here does show that this metric is reasonably robust to the variation of other parameters, i.e., the shape of the gradient is sufficiently distinct to allow us to characterize the scale radius of the dark matter halo.

We have shown the phase behavior here at the time when a given simulation best fits the Fourier amplitudes of the HI data.  CB09 and CBCB have outlined how the best-fit simulation from a simulation parameter survey can be displayed on a variance vs variance plot.  Disturbances in the gas disk dissipate on the order of a dynamical time.  As such, even the best-fit simulation traces a trajectory on the variance vs variance plot, with the minimum yielding the best-fit time.  We have examined the time variation of these simulations as well.  We do not find that the time variation of the phase would allow distinct scale radius cases to mimic each other.  We conclude therefore that the phase of the $m=1$ mode allows us to discriminate between different scale radii within a factor of $\sim 1.5$ (i.e., we can clearly discriminate the $R_{s}=17$ case from the $R_{s}=11$ case).   

\section{Discussion}
There are some caveats in our work here that are worth noting for the application of this method in generality.  We have not investigated in this paper the effect of non-spherical halos on either the Fourier amplitudes or the phase.  Recent work has studied the effects of prolate halos (Banerjee \& Jog 2011) on the vertical structure of the Milky Way's HI disk.  We defer a detailed investigation of these effects to a future paper.  Preliminary work (Chakrabarti, Debattista \& Blitz, in preparation) finds that the evolution of the halo shape (and thereby the strength of the Fourier amplitudes in the outskirts of the gas disk as determined by intrinsic processes) is significantly affected by gas cooling and angular momentum transport from the gas to the halo.  Debattista et al. (2008) found earlier that the presence of a baryonic (stellar) disk affected the shape of the dark matter halo, leading to a rounder halo as the disk is grown.  We have recently examined similar simulations which include a gaseous component (Chakrabarti, Debattista \& Blitz, in prep) and find that the presence of the gaseous component renders halo shapes more spherical over time after the onset of gas cooling.  The shapes of halos are considerably rounder close to present day, when the stellar to dark matter masses are comparable to local spirals ($M_{\star}/M_{\rm DM} \sim 0.03$) (Leroy et al. 2008), relative to $z \sim 2$.  Examination of the Fourier amplitudes of these simulations close to present day indicates that the Fourier amplitudes in the outskirts are $\la$ 10 \%, which suggests that non-spherical halos would not be the primary contributor to the observed strength of the Fourier amplitudes in local spirals.  This is certainly valid for M51, which is known to be a tidally interacting system (CBCB; Salo \& Laurikainen 2000).  Local spirals that display small ($\la$ 10 \% in the Fourier amplitudes, as defined in CB09, relative to the axisymmetric mode) perturbations in the outskirts, and have no visible tidally interacting companion may require a more careful treatment of the effect of halo shapes.  Nonetheless, the presence of a cold disssipative gaseous component needs to be modeled to more comprehensively understand the evolution of halo shapes.  While cosmological simulations (Maccio et al. 2008) find halo shapes rendered non-spherical (in many cases triaxial) by repeated mergers, this effect may be mitigated in reality by a cold gaseous component transferring its abundance of angular momentum to the dark halo, thereby rendering it more spherical.

Our observationally motivated inference of the scale radius of the dark matter halo may be useful for a number of reasons.  Firstly, the NFW density profile has some dependence on the spectral index of primordial density fluctuations (NFW96; NFW97).   Secondly, this profile emerges within the context of dissipationless, cosmological simulations, and baryonic effects may well serve to alter this profile.  Therefore, an independent observationally motivated probe of the density profile for specific spiral galaxies can serve to support the theoretical bases of this so-called universal density profile, wherein galaxies have grown hierarchically from primordial density fluctuations.   Performing such a study for a large sample of spirals in the Local Volume, as can be done in the near future with galaxies from the THINGS survey, will allow for a statistical determination of the scale radii of spiral galaxies.  Such a study can be performed to zeroth-order using the scaling relations in Chang \& Chakrabarti (2011) that allow one to infer the satellite mass from the observed HI map, and refined with numerical calculation.  Another important tracer of the mass distribution is the observed rotation curve (Bosma 1978; Rubin et al. 1982; Blitz 1979).  The rotation curve for many spirals cannot be determined beyond $\sim$ 10 disk scale lengths.  We leave a comparison of this method to the rotation curve to a future paper.  It would also be worthwhile to compare our approach to gravitational lensing (Treu \& Koopmans 2002; Mandelbaum et al. 2006), which also serves to probe mass distributions independent of the stellar light.  Our method is ideally suited for analysis of disturbances in the outskirts of spiral galaxies in the Local Volume, and can in the future be applied to a higher redshift population with observations from instruments such as SKA.

\section{Conclusion}

$\bullet$  We find that the shape of disturbances in the fragile, extended HI disks of galaxies reflects the underlying density profile of the dark matter halo in spiral galaxies.  We employ the phase of the $m=1$ mode as our primary metric of comparison to observed HI data.  Building on prior results from CBCB where the satellite mass and pericentric approach distance was determined for M51, we find that the scale radius of the dark matter halo can be determined from the phase of the $m=1$ mode.  

$\bullet$ The three regimes of interest for the phase of the $m=1$ mode that delineate the variation of the scale radius and how it is reflected in the phase are:  $d/dr(\phi_{1})<0$ for $r/R_{s} < 1$, a transitional region at $r /R_{s}\sim 1$, and $d/dr(\phi_{1})>0$ for $r/R_{s} >1$ til the edge of the gas disk.   Long baseline HI observations that can reveal this transitional region between the $r^{-1}$ shallow profile (that would produce more loosely wrapped spirals) and the steeper $r^{-3}$ profile (that would produce more tightly wrapped spirals) will allow us to quantitatively infer the scale radius of dark matter halos of $\it{specific~spiral~galaxies}$.  The slight uncertainty in the phase that arises due to variations in initial conditions and orbits does not preclude us from inferring the value of the scale radius to within a factor of $\sim 1.5$.  We have discussed preliminary work that suggests that other effects (such as intrinsic processes arising from non-spherical halos) will not significantly affect our results here.  

$\bullet$ Application of this method to the Whirpool Galaxy yields a scale radius of $17~\rm kpc$, which is consistent with the expected range of scale radii of $\sim 10^{12} M_{\odot}$ halos in dissipationless cosmological simulations.

$\bullet$  We have also demonstrated that for $r/R_{s} > 1$, our results depend essentially on the scale radius and not on the concentration parameter.  These results suggest that we can utilize observed disturbances in the extended HI disks of galaxies to characterize the density profile of the dark matter halo in spiral galaxies.  In the future, we will apply this method to a large sample of spirals to obtain a statistical determination of the scale radii of spiral galaxies in the Local Volume.

\bigskip
\bigskip

\acknowledgements
We thank Lars Hernquist, Chris McKee, Phil Chang, Pedro Marronetti, Leo Blitz, Jay Gallagher, Risa Wechsler, and Daniel Holz for helpful discussions.  

\newpage                                                                          
\references 

Banerjee, A. \& Jog, C., 2011, ApJL, 732L, 8B \\
Bigiel, F., Leroy, A., Seibert, M., et al., 2010, {\it Astronomical Journal}, 140, 1194 \\
Blitz, L., ApJ, 1979, 231, L115 \\
Bullock, J., Dekel, A., Kolatt, T.S., et al., 2001, ApJ, 555, 240 \\
Chakrabarti, S. \& Blitz L., 2009, {\it \mnras\/}, 399, L118 [CB09]  \\
Chakrabarti, S. \& Blitz L., 2011, {\it \apj\/}, 731, 40C  \\
Chang,P. \& Chakrabarti, S., 2011, MNRAS, 416, 618C \\
Chakrabarti, Bigiel, Chang \& Blitz 2011, ApJ,  743, 35 [CBCB] \\
Debattista, V., Moore, B., Quinn, T., et al., 2004, ApJ, 681, 1076 \\
Dobbs, C.L., et al.2010,  {\it \mnras\/} {\bf 403}, 625 \\
Dubinski, J., Mihos, C. \& Hernquist, L., 1996, ApJ, 462, 576 \\
Gnedin, O., Kravtsov, A., Klypin, A. \& Nagai, D., 2004, ApJ, 616, 16 \\
Hernquist 1990, ApJ, 356, 359H \\
Mihos, C., Dubinski, J. \& Hernquist, L., 1998, ApJ, 494, 183 \\
Maccio, A. et al., 2008,  {\it \mnras\/} {\bf 391}, 1940 \\
Navarro, J., Frenk, C.S. \& White, S.D.M., 1996, ApJ, 462, 563 [NFW96] \\
Rubin, V., Thonnard, N. \& Ford, K.W., 1977, ApJL, 217, L1 \\
Salo, H. \& Laurikainen, E., 2000,  {\it \mnras\/} {\bf 319} \\
Smith, J.,  et al., 1990, {\it \apj\/} {\bf 362}, 455S \\
Springel, V., 2005,  {\it \mnras\/} {\bf 364}, 1105 \\
Springel, V., Frenk, C.S., \& White, S.D.M., 2006, Nature, 440, 1137 \\
Thilker, D., Bianchi, L., Meurer, G., et al, 2007, ApJS, 173, 538 \\
Treu, T. \& Koopmans, L., 2002, ApJ, 575, 84 \\
Walter, F. et al., 2008, {\it Astronomical Journal} {\bf 136}, 2563 \\

\end{document}